\documentclass[pdflatex,sn-mathphys-num]{sn-jnl}

\usepackage{graphicx}%
\usepackage{multirow}%
\usepackage{amsmath,amssymb,amsfonts}%
\usepackage{amsthm}%
\usepackage{mathrsfs}%
\usepackage[title]{appendix}%
\usepackage{xcolor}%
\usepackage{textcomp}%
\usepackage{manyfoot}%
\usepackage{booktabs}%
\usepackage{algorithm}%
\usepackage{algorithmicx}%
\usepackage{algpseudocode}%
\usepackage{listings}%
\usepackage{listings}
\usepackage{color}
\usepackage{xcolor}
\usepackage{graphicx}

\definecolor{dkblue}{rgb}{0,0.39,0}
\definecolor{gray}{rgb}{0.66,0.66,0.66}
\definecolor{mauve}{rgb}{0.91,0.33,0.50}
\definecolor{gold}{rgb}{1,0.84,0}

\begin{document}

\title[Exploring LLMs for Malware Detection: Review, Framework Design, and Countermeasure Approaches]{Exploring LLMs for Malware Detection: Review, Framework Design, and Countermeasure Approaches}

\author[1,2]{\fnm{Jamal} \sur{Al-Karaki}}\email{Jamal.Al-Karaki@zu.ac.ae}

\author*[1]{\fnm{Muhammad Al-Zafar} \sur{Khan}}\email{Muhammad.Al-ZafarKhan@zu.ac.ae}

\author[3]{\fnm{Marwan} \sur{Omar (Member, IEEE)}}\email{momar3@iit.edu}

\affil*[1]{\orgname{College of Interdisciplinary Studies (CIS), Zayed University}, \state{Abu Dhabi}, \country{UAE}}

\affil[2]{\orgdiv{College of Engineering}, \orgname{The Hashemite University}, \city{Zarqa}, \country{Jordan}}

\affil[3]{\orgdiv{Department of Information Technology and Management}, \orgname{Illinois Institute of Technology}, \city{Chicago, IL}, \country{USA}}

\abstract{\footnotesize{The rising use of Large Language Models (LLMs) to create and disseminate malware poses a significant cybersecurity challenge due to their ability to generate and distribute attacks with ease. A single prompt can initiate a wide array of malicious activities. This paper addresses this critical issue through a multifaceted approach. First, we provide a comprehensive overview of LLMs and their role in malware detection from diverse sources. We examine five specific applications of LLMs: Malware honeypots, identification of text-based threats, code analysis for detecting malicious intent, trend analysis of malware, and detection of non-standard disguised malware. Our review includes a detailed analysis of the existing literature and establishes guiding principles for the secure use of LLMs. We also introduce a classification scheme to categorize the relevant literature. Second, we propose performance metrics to assess the effectiveness of LLMs in these contexts. Third, we present a risk mitigation framework designed to prevent malware by leveraging LLMs. Finally, we evaluate the performance of our proposed risk mitigation strategies against various factors and demonstrate their effectiveness in countering LLM-enabled malware. The paper concludes by suggesting future advancements and areas requiring deeper exploration in this fascinating field of artificial intelligence.}}


\maketitle
\section{Introduction}

The swift evolution of malware threats poses significant challenges to cybersecurity. Conventional malware detection techniques often struggle to keep up with the sophisticated and polymorphic malware that continually adapts to evade detection. Recent advancements in Natural Language Processing (NLP), particularly through the use of Large Language Models (LLMs), present a promising new approach to addressing these challenges. LLMs, with their advanced capabilities in understanding and generating human-like text, have demonstrated exceptional performance across various domains, including text classification and anomaly detection. This paper explores the potential of leveraging LLMs for malware detection via articulating a systematic review of the state-of-the-art in the field, focusing on their ability to analyze and interpret complex patterns in executable code and network traffic. By harnessing the power of LLMs, we aim to enhance and shed light on the in-vogue practices that exist and characterize the various literature pieces that exist in a consolidated and sequential format, offering a novel perspective on combating the evolving threat landscape.

In particular, a comprehensive introduction to LLMs and malware and how, in general, LLMs can be used to detect malware attacks since its advent in November 2022, ChatGPT,  released by OpenAI \cite{roumeliotis2023chatgpt,wu2023brief,deng2022benefits}, has been used by consumers from all backgrounds and for various domains. Subsequently, many other big tech companies have pooled their resources and strategically directed their research teams to create and put into production their own versions of LLMs. In general, LLMs work as follows: 
\begin{enumerate}
\item \textbf{Tokenization:} Input text (prompts) are broken down into small units called \textit{tokens} and assigned a unique integer value that is used for identification of position in the LLM's underlying model. Essentially, there are three primary tokenization methods: For words, for subwords, and for characters. 

Given the input prompt ``This is my prompt'', one potential tokenization is
\begin{align}
\begin{aligned}
\mathcal{T}\left(\text{``This is my prompt''}\right)\longrightarrow\left[\text{``This''}, \text{``is''}, \text{``my''}, \text{``prompt''}\right].
\end{aligned}
\end{align}

\item \textbf{Encoding:} Tokens are converted to numerical embedding vectors that capture semantic information about the tokens. This facilitates the model's understanding of the contextual nature of the tokenized words.

A potential encoding of the above tokens is
\begin{equation}
\mathcal{E}\left[\text{``This''}, \text{``is''}, \text{``my''}, \text{``prompt''}\right]\longrightarrow\left[0,1,2,3\right].
\end{equation}

\item \textbf{Positional Encoding:} In order to incorporate positional information of the numerical embeddings, positional encoding is performed in order to provide the underlying LLM model with information about the order of the tokens in the input sequence. This facilitates the differentiation between tokens based on their relative dispositions. Mathematically, positional encoding is given by
\begin{align}
\begin{aligned}
\mathcal{PE}(\langle i,j\rangle,2k)=&\;\sin\left(\frac{\langle i,j\rangle}{10^{8k/d}}\right), \\
\mathcal{PE}(\langle i,j\rangle,2k+1)=&\;\cos\left(\frac{\langle i,j\rangle}{10^{8k/d}}\right),
\end{aligned}
\end{align}
where $\langle i,j\rangle$ is the position of the token in the sequence, $k$ is the dimension index of the embedding vector, and $d$ is the dimensionality of the token embeddings. 
\item \textbf{Transformer Layer:} The transformer layer is composed of feedforward neural network layers that use the \textit{self-attention mechanism} to compare the input sequences to each other in a hierarchical manner. This allows for context to be captured effectively, and the encapsulation of long-range dependencies. Mathematically, the self-attention mechanism is given by
\begin{equation}
\mathcal{A}(\mathbf{Q},\mathbf{K},\mathbf{V})=\sigma\left(\frac{\mathbf{QK}^{T}}{\sqrt{d_{k}}}\right)\mathbf{V},
\end{equation}
where $\mathbf{Q}$ is the query matrix, $\mathbf{K}$ is the key matrix, $d_{k}$ is the dimensionality of the key vector, $\mathbf{V}$ is the value matrix, and $\sigma(\xi_{i})=\exp\left(\xi_{i}\right)/\sum_{j}\exp\left(\xi_{j}\right)$ is the softmax activation function, with $\xi$ being a placeholder. 

The output from the attention mechanism is then fed into the feedforward neural network, which applies a nonlinear transformation to the encoded tokens in order to develop complex relationships that are not possible with the assumption of simple linear relationships; commonly, this is chosen to be the ReLU function. Mathematically, the application of the feedforward neural network yields
\begin{equation}
\mathcal{F}(\mathbf{x})=\sum_{\ell=1}^{L-1}\text{ReLU}\left(\mathbf{x}^{T}\mathbf{W}_{\ell}+\mathbf{b}_{\ell}\right)\mathbf{W}_{\ell+1}+\mathbf{b}_{\ell+1},
\end{equation}
where $\mathbf{x}$ is the tokenized input embedding, $\mathbf{W}_{\ell}$ are the weights in the $\ell^{\text{th}}$ layer, $\mathbf{b}_{\ell}$ are the bias terms in the $\ell^{\text{th}}$ layer, $L$ is the number of layers, and $\text{ReLU}(\eta)=\max(0,\eta)$ is the ReLU activation function, with $\eta$ being a placeholder. 

\begin{figure*}[htpb]
\centering
\includegraphics[width=1.05\linewidth]{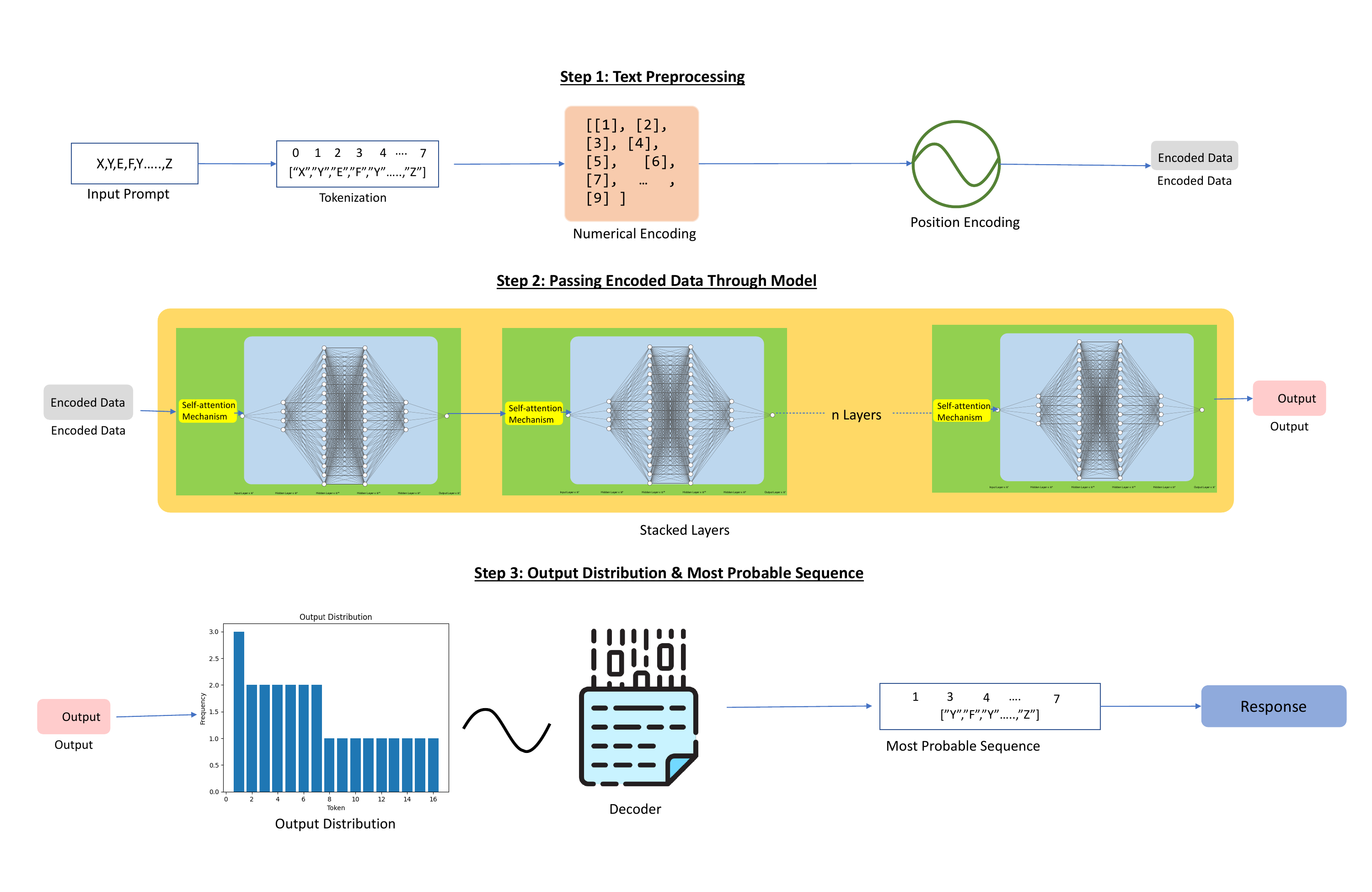}
\caption{A simplified overview of a Large Language Model (LLM) pipeline. While it is a gross oversimplification of the intricate architecture and operations of a real LLM, it effectively illustrates the core steps involved in generating text responses, namely: Tokenization, encoding, positional encoding, various stacked layers, the attention mechanism, the output distribution, and the most probable sequence generation.}
\label{fig1}
\end{figure*}

The steps $1-5$ are summarized in the architecture diagram Fig. \ref{fig1}. 

An important feature of LLMs is that they contain multiple transformer layers stacked. The output from one layer becomes the input to the next layer. The purpose of stacking is for the refinement of the tokens, and to increasingly capture complex patterns from the data.
\item \textbf{Decoding:} Upon passing through the stacked transformer layers and undergoing refinement, the decoder samples from the output distribution or uses a heuristic search algorithm like beam search to select the most probable sequence of tokens. 
\end{enumerate}

Since the launch of ChatGPT, every aspect of human life has been affected. From education to internet search, to science, law, medicine, and even research, there has been a rise in people using it as a tool to achieve their objectives, be it for casual use or in their work. The latest version, as of March 2024, powered by the GPT-4 engine, has been demonstrated to perform substantially well in major benchmarking examinations, consistently scoring in the upper percentiles, like the New York Bar Exam (Law) \cite{katz2024gpt}, the SAT exam for aspirant US-college entrants, various Advanced Placement (AP) subject examinations \cite{achiam2023gpt}, orthopedic in-training examinations \cite{kung2023evaluating}, United States Medical Licensing Exam (USMLE) \cite{gilson2023does,epstein2023variability}, the University of Pennsylvania's Wharton School of Business MBA exam \cite{terwiesch2023would1,terwiesch2023would2}, a radiology-style board exam \cite{bhayana2023performance}, and is currently being tested out to determine its performance on other standardized examinations that are notoriously difficult; a more comprehensive discussion of multiple choice-type examinations passed by ChatGPT is provided in \cite{newton2023chatgpt}.

Additionally, many businesses have been built on LLMs; for example, there are sites that offer to build entire presentation decks on any topic or generative models that can generate images and video based on text prompts and an overabundance of other applications; see, for example, \cite{short2023artificially} for the business impact. This has led to a division in society and amongst experts as to whether it's something good to have in the first place, and more especially to have them accessible to members of society engaged in foundational and critical tasks like children at school, as the arguments posited is that it leads to children not being innovative and stretching their minds and imaginations to think critically and originally. 

A characterization of how ``powerful'' an LLM is, is by the number of parameters that it has. Table~\ref{tab1} tabulates some of the most common examples of LLMs.

\begin{table}[h]
\centering
\begin{tabular}{p{3cm}p{5cm}p{4cm}}
\hline
\textbf{LLM} &\textbf{Parent Company} &\textbf{Number of Parameters} \\
\hline
ChatGPT &OpenAI &GPT-3: $175\times 10^{9}$ \\
        & &GPT-4: $1.76\times 10^{12}$   \\
Gemini / Bard  &DeepMind &$30-65\times 10^{12}$ \\
Falcon &Technology Innovation Institute &$180\times 10^{9}$ \\ 
BloombergGPT &Bloomberg &$50\times 10^{9}$ \\
LLaMa &Meta &$65\times 10^{9}$ \\
Grok &X.ai &$314\times 10^{9}$ \\
BERT &Google &$345\times 10^{6}$ \\
PaLM &Google &$540\times 10^{9}$ \\
Orca &Microsoft &$13\times 10^{9}$ \\
Claude &Anthropic &$137\times 10^{9}$ \\
\hline 
\end{tabular}
\caption{Table 1: Benchmarking of some popular LLMs.}
\label{tab1}
\end{table}

Originating from ideas in NLP, nowadays, LLMs use Deep Learning (DL) models, like transformers, to find statistical relationships between words and phrases in its training data/corpus. While LLMs can seemingly answer any question, they oftentimes suffer from the hallucination problem \cite{zhang2023siren,rawte2023survey,huang2023survey} in which they make up information that seems correct but is completely fallacious. While unlocking nearly infinite possibilities in their ability to answer questions on the go, LLMs are not Artificial General Intelligence (AGI) and, thus, cannot autonomously think and make decisions for themselves and, therefore, require prompting. 

\textit{malware}, the portmanteau of the words ``malicious'' and ``software'' is a software program that is specifically designed to cause harm to a computer system/network. Hackers have used malware to disrupt operations in a computer system, using the software to overwork the hardware and therefore causing damage to the hardware, stealing data, and even spying, and is therefore viewed as a major threat to society, governments, and can potentially cause global disorder due to the financial implications, and the leaking of sensitive information caused by data breaches. Computing security experts have categorized malware into the following groupings:
\begin{enumerate}
\item \textbf{Worms:} Hostile self-replicating programs that spread throughout the computing system without attaching to other programs.
\item \textbf{Spyware:} Using to steal critical information. These include passwords, personal data, login credentials for online banking, and so on.
\item \textbf{Ransomware:} Programs that restrict, or completely block, access to a computing system. Oftentimes, they block access to certain files, and the hacker behind its deployment demands monetary rewards (cash or untraceable cryptocurrencies) in order to restore access.
\item \textbf{Viruses:} Programs that replicate themselves and maliciously couple to other programs with the goal of spreading and infecting the computer as they are being used.
\item \textbf{Trojan Horses:} Programs that disguise themselves as legitimate programs from reputable companies, with the intention of being accessed in order to cause malicious damage to its host.  
\end{enumerate}

In Table \ref{tab2}, we juxtapose how various LLMs fair in comparison to one another for the task of malware detection.

\begin{table}[htpb]
\centering
\begin{tabular}{p{2cm}p{10cm}}
\hline
\textbf{LLM} &\textbf{Performance} \\
\hline
 BERT &$\bullet$ Malware samples are represented by a sequence of tokens. \\
  &$\bullet$ Contextual information is encapsulated effectively due to its bidirectional nature. \\
  &$\bullet$ BERT shows superior performance when detecting subtle malware patterns. \\
\hline 
RoBERTa &$\bullet$ An optimized version of BERT that addresses some of BERT's limitations when it comes to NLP tasks. \\
 &$\bullet$ Potentially offers advanced performance when compared with BERT due to the robust training methodology incorporated in its architecture. \\
\hline
AlBERTa &$\bullet$ A compact/lite variation of BERT that achieves analogous performance but has fewer parameters. \\ 
 &$\bullet$ Highly applicable for detecting malware in resource-constrained environments. This is because it uses less computational resources. \\
\hline
GPT-2 &$\bullet$ These LLM models are trained on a large corpus of text data. \\
and GPT-3 &$\bullet$ Due to the rigorous training process, these models can generate features for subsequent analysis or be fine-tuned to detect malware effectively. \\
\hline 
 XLNet$^{*}$ &$\bullet$ Not very often used for malware detection tasks. \\
  &$\bullet$ Due to its advanced generalized autoregressive pretraining, it outperforms BERT and its variants on many other NLP tasks. Thus, it can potentially be used for enhanced performance. \\
\hline 
\end{tabular}
\caption{A comparison of various LLM models for malware detection.}
\label{tab2}
\end{table}

The foremost objective of this paper is to discuss, in sufficient detail, how LLMs can be used to detect camouflaged patterns that may exist within code that may be nearly impossible for an expert system -- like a human -- to detect. We are conscious of the fact that the construction of a survey paper in this field is not novel; however, the purpose of this paper is to give accounts of the latest developments in the field -- Particularly how LLMs are being used for malware detection, address various shortcomings of the current frameworks that exist in order to provide an attentions of these risks and target a broader scope of security researchers interested in the field. The novel contributions of this paper are:
\begin{enumerate}
\item A comprehensive exploration of the various literature pieces in the field.
\item A propoundment of a set of guiding principles to steer LLMs in the correct direction for detecting malware.
\item The proposal of a risk mitigation framework on leveraging LLMs for detecting malware from software. 
\item A discussion on the \textit{faux pas} related to using LLMs for detecting malware from software. 
\end{enumerate}

The rest of the paper is organized as follows:
\begin{itemize}
\item In Sec. \ref{literature review}, we provide an organized and exhaustive coverage of the research in the field.
\item In Sec. \ref{problem formulation}, we use our understanding from the literature to formulate metrics for using LLMs as a tool to prevent different attack types. 
\item In Sec. \ref{guidelines}, we propose a set of guidelines and risk mitigation strategies for preventing malware attacks using LLMs.
\item In Sec. \ref{performance evaluation}, we demonstrate the use of the metrics from Sec. \ref{problem formulation} with fictitious data that has the potential to be collected in the real world.
\item In Sec. \ref{conclusion}, we conclude on the findings of this paper and provide a discussion on why LLMs cannot possibly prevent all malware attacks. 
\end{itemize}


\section{Literature Review} \label{literature review}
In this section, we peruse the various literature sources relevant to the core objective of using LLMs for malware detection.

In order to classify the literature into buckets with common trends, we use the following scheme to denote the various characterizations described in Tab. \ref{tab200}.

\begin{table}[h]
\centering
\begin{tabular}{|p{3cm}|p{4cm}|p{5cm}|}
\hline 
\textbf{Parameter/Symbol} &\textbf{Definition} &\textbf{Description} \\
\hline
$\textcolor{red}{\alpha}$ &Using LLMs for harmful content generation. &Refers to those papers where LLMs are used for creating content that is detrimental. \\
\hline
$\textcolor{red}{\beta}$ &Exploitation of LLMs for malicious weaponization. &Those papers where LLMs have been, or have the potential to, be utilized for militarization. \\
\hline
$\textcolor{red}{\gamma}$ &The dissemination of malware through, or by the usage of, LLMs. &Those papers that use, or have the potential to use, LLMs for the distribution of malware. \\
\hline
$\textcolor{red}{\delta}$ &Survey papers of the field. &Those research papers that provide a review of the existing research in the field of LLMs for malware detection. \\
\hline
$\textcolor{red}{\epsilon}$ &Benchmarking LLMs on malware. &Those papers that test out various LLMs on different malware in order to provide a gauge. \\
\hline
$\textcolor{red}{\zeta}$ &Proposal of a new and effective model. &Those papers that advance new LLMs for malware detection. \\
\hline 
\textcolor{red}{*} &Out of the ordinary paper, but still relevant. &Those papers that are atypical of the research in the field of LLMs for malware detection. \\
\hline 
\end{tabular}
\caption{Description of the various symbols used for characterizing the different literature papers.}
\label{tab200}
\end{table}

Below, we provide a review of the most prominent pieces of work on the subject matter; this comprises survey papers and contemporary works in the field.

In\cite{yao2024survey} (\textcolor{red}{$\delta$}), the authors explore the positive impact of LLMs on security and privacy, prospective threats, and the susceptibilities of LLMs. This work characterizes the various papers into three types: Those with beneficial applications, those with offensive applications, and those having vulnerabilities and defenses.  

In \cite{beckerich2023ratgpt} (\textcolor{red}{$\alpha$}, \textcolor{red}{$\beta$}, \textcolor{red}{$\gamma$}), the authors present a proof-of-concept that illustrates the usage of ChatGPT as an agent for delivering malware while evading detection. With a particular emphasis on command-and-control (C2) servers, the authors underscore how to leverage LLMs for executing commands on the victim's computer. Lastly, the paper explores the need for risk mitigation strategies, frameworks, and guidelines that need to be developed and adopted as a preventative measure. 

In \cite{shayegani2023survey} (\textcolor{red}{$\delta$}), the authors focus on the intersection of NLP and security, with a specific emphasis on adversarial attacks on nefarious attacks that can mislead LLMs. The paper characterizes the literature papers into various attack types, namely: Textual attacks, multi-modal attacks, and attacks on sophisticated systems like federated and self-organizing (multi-agent) systems. This paper furnishes an examination of various attack types and makes the field more accessible to researchers and laypeople unfamiliar with the topic. 

In \cite{motlagh2024large} (\textcolor{red}{$\delta$}), the authors focus on the defensive and antipathetic use of LLMs in cybersecurity and present a SWOT-like analysis of the state-of-the-art, with a particular focus on understanding the extent of risks and opportunities. The added value of this research is the identification of research gaps like limited defensive applications, scalability, robustness, explainability, data privacy and ethical considerations, and adoption.  

In \cite{xu2024autoattacker} (\textcolor{red}{$\alpha$}), the authors focus on simulating human-aided post-breach attacks on LLMs. The richness of this research lies in its scope and focus; whereas other papers in the field focus on the pre-breach (such as malware and phishing) phase of an attack, the sole focus of this research is after a breach occurs. 

\begin{figure*}[htpb]
\centering
\includegraphics[width=1\linewidth]{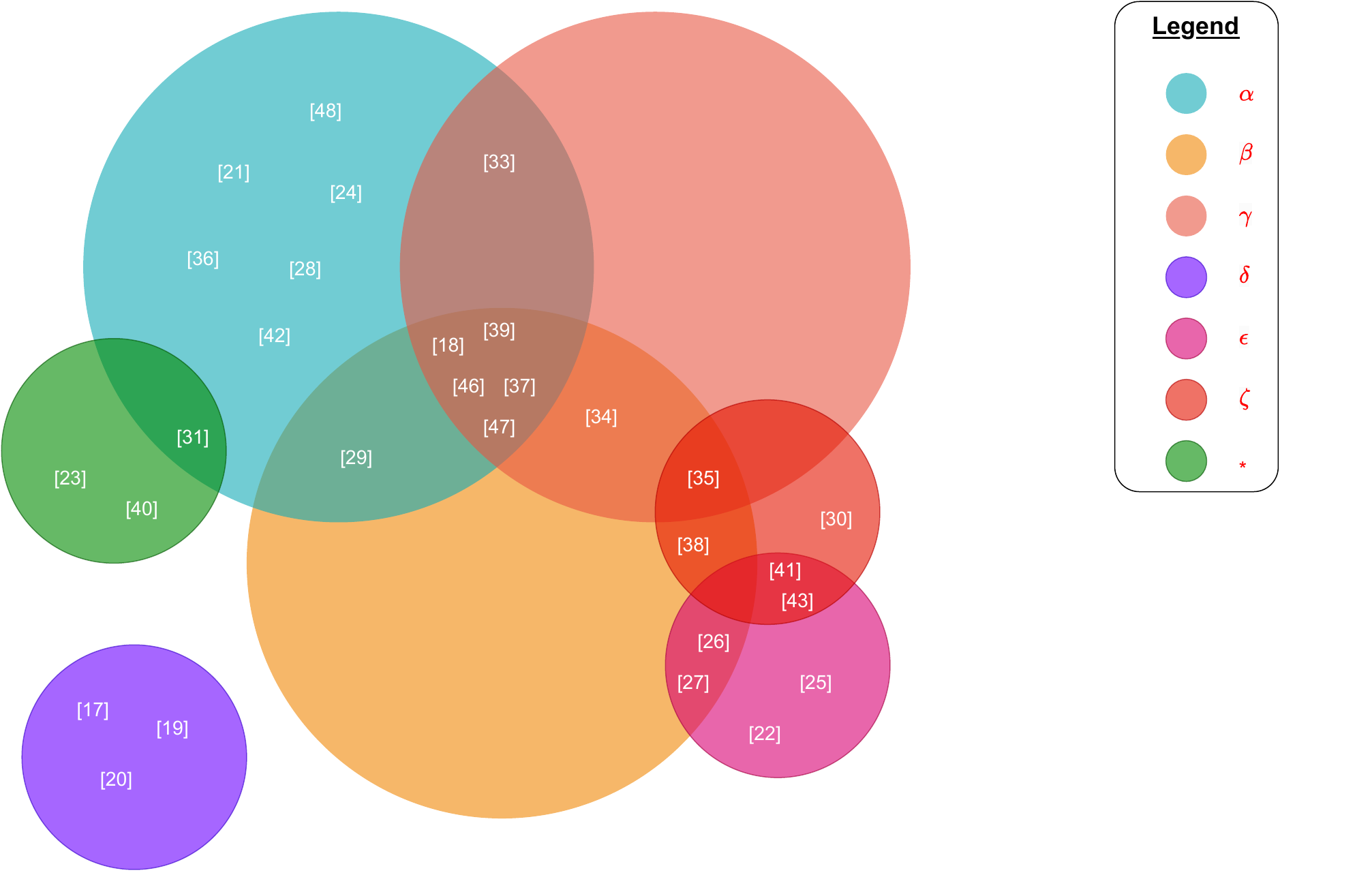}
\caption{Visual taxonomy of the overlapping categories of classifications within the literature base. Each colored circle represents a distinct category, and the overlapping regions indicate the shared literature pieces between these categories.}
\label{fig3}
\end{figure*}

In \cite{vasconcelos2023llama} (\textcolor{red}{$\epsilon$}), the authors use two production-grade LLMs, LLaMa-12B, and IDA Pro, to investigate the strategic shift from crypto-ransomware to data-theft ransomware. Specifically, the two LLMs in question are used to emulate and analyze ransomware variants. The findings underscored the necessity for adaptive cybersecurity strategies incorporating advanced detection systems to recognize ransomware activities effectively.

In \cite{agrawal2024review} (\textcolor{red}{*}), while strictly not a paper focused on LLMs for malware detection, focuses on the generation of synthetic/inorganic malware-based data using Generative Adversarial Attacks (GANs). The importance of this work cannot be overstressed as data that shows how LLMs can be used to detect malware in code is scant and not widely available because of the exposure of personalized intellectual property (code) of organizations. The research shows that GANs are very effective in generating realistic cyberattacks, and thereafter Deep Learning (DL) models are effective in their classifications. This work lays the foundation for the development of algorithms for using LLMs to detect malware attacks.   

In \cite{yamin2024applications} (\textcolor{red}{$\alpha$}), the authors use Turing exploration into ML for generating complex and evolving cybersecurity scenarios for LLMs to detect by exploiting their generative ability. This research focuses on the transformation of the impediment of the hallucination problem of LLMs into an advantage for creating new scenarios for the LLM to train on. 

In \cite{li2023efficient} (\textcolor{red}{$\epsilon$}), the authors use LLaMa-7B to test the proof-of-concept of a novel ransomware detection methodology that involves the analysis of portable executable (PE) files. Specifically, the procedure involves the conversion of the PE files into gray-scale bitmap images and then examining them using LLaMa-7B. The highlight of this research was the approach adopted. 

In \cite{greshake2023not} (\textcolor{red}{$\beta$}, \textcolor{red}{$\epsilon$}), the authors focus on the identification of the risks associated with indirect prompt injection attacks on LLMs. The research highlights that strategic prompting can influence the behavior of LLMs and remotely capitalize on LLM predisposition. The novelties of this research lie in its overarching gradation of attack vectors: Worming, data theft, ecosystem contamination, etc., and by the viability of real-world applications using GPT-4 (on synthetic data) and prompt injection on Bing Chat and code-completion engines (organic data). Similar findings were established in \cite{liu2023prompt}. 

In \cite{kang2023exploiting} (\textcolor{red}{$\alpha$}), the authors explore how LLMs can be used for both positive and negative activities; the \textit{dual-use risk}. The study demonstrates how instruction-following LLMs can produce targeted malicious content (spam, hate speech, etc.) and get around defenses set up on LLM APIs. The findings underscore the need for innovative approaches to preventing and neutralizing the risks posed by instruction-following LLMs, and addressing the challenges presented by increasingly sophisticated adversaries and attacks.

In \cite{madani2023metamorphic} (\textcolor{red}{$\alpha$}, \textcolor{red}{$\beta$}), the author focuses on the usage and potential of LLMs for the creation of metamorphic malware -- Malware that automatically adjusts its code for evading detection. The paper introduced a framework, based on LLM architecture, for creating programs that test themselves for mutation. The paper is also future-focusing in that it focuses on the next generation of malware detection.

In \cite{devadiga2023gleam} (\textcolor{red}{$\zeta$}), the authors address a primary concern of malware detection: How to detect it when it continuously evolves to evade identification. By combining hex- and op-code features with LLMs, they propose a new model that generates synthetic samples reminiscent of evasive malware. The experimental results indicate that the model showed substantial performance. However, further improvements are required to achieve higher. The novelty of the research, besides the new model, is the particular focus on evasion tactics like obfuscation and encryption. 

In \cite{glukhov2023llm} (\textcolor{red}{$\alpha$}, \textcolor{red}{*}), the authors focus on the challenges and risks associated with LLMs being used for generating malicious content. The research reasons that semantic censorship needs to be reexamined in the context of being a security threat. Admittedly, this paper has indirect relevance because it helps influence the preventative principles that are posited in  Sec. \ref{risk mitigation}.

In \cite{fujima2023semantic} (\textcolor{red}{$\epsilon$}), the authors use an ML model powered by ChatGPT to detect phishing emails. By analyzing real-world phishing emails, the research identifies psychological tactics, technical deception, and language anomalies that are habitually used to plant ransomware within organizations.

In \cite{pa2023attacker} (\textcolor{red}{$\alpha$}, \textcolor{red}{$\gamma$}), the authors focus on the generation of malware using ChatGPT, text-davinci-003, and AutoGPT. Using approximately 400 lines of code, seven malware programs, and two attack tools within a short timeframe of 90 minutes, including debugging time, were developed. This demonstrates that within such a short time period, and within the wrong hands, the destructive capability of these models, once again highlighting the dual-use risk discussed in \cite{kang2023exploiting}. Lastly, the authors provide discernment into the current barriers and areas for improvement in AI safety mechanisms in LLMs.

In \cite{alawida2024unveiling} (\textcolor{red}{$\beta$}, \textcolor{red}{$\gamma$}), the foremost objective of the authors is to determine the relationship between ChatGPT and cyberattacks. In particular, they probe how malicious actors exploit ChatGPT to launch various cyberattacks. It was found that malicious actors posed as regular users to manipulate the model's vulnerability to hostility. The novelties of this research include the identification of tactics employed by malicious actors and an investigation into user perceptions, with the goal of identifying the weaknesses and biases and proposing mitigation strategies. 

In \cite{begou2023exploring} (\textcolor{red}{$\beta$}, \textcolor{red}{$\gamma$}, \textcolor{red}{$\zeta$}), the authors investigate how ChatGPT can be used to automate various stages of a phishing attack, and automatically deploy malware to execute these attacks. The novelties of this research include: Generating an automated phishing kit and raising awareness about LLM misuse.   

In \cite{fang2024llm} (\textcolor{red}{$\alpha$}), the authors focus on using LLMs to hack websites. The research identifies the empowering strengths that LLMs have that facilitate the hacking process: The ability to read documents, interact with tools, and recursively call themselves. The paper argues that frontier models, like GPT-4, through their advanced features and adeptness, can drive cybersecurity capabilities. 

In \cite{gupta2023chatgpt} (\textcolor{red}{$\alpha$}, \textcolor{red}{$\beta$}, \textcolor{red}{$\gamma$}), the authors examine the evolution of GenAI in the context of cybersecurity. The research demonstrates attacks such as jailbreaks, reverse psychology, and prompt injection on ChatGPT. In particular, how LLMs and general GenAI tools develop various cyber attacks centering around social engineering, phishing, automated hacking, attack payload generation, malware creation, and polymorphic malware. This research exposes ChatGPT's vulnerabilities and the presentation of defense techniques. 

In \cite{ferrag2023revolutionizing} (\textcolor{red}{$\beta$}, \textcolor{red}{$\zeta$}), the authors focus on using BERT for cyber threat detection in IoT networks. Specifically, a new model security based upon the BERT architecture is presented for detecting cyber threats, with the usage of a novel privacy-preserving encoding technique. It should be borne in mind that technically speaking, BERT is a Small Language Model (SLM); however, the methods developed and employed are scalable to LLMs. The model is demonstrated to have high performance in threat detection. 

In \cite{charan2023text} (\textcolor{red}{$\alpha$}, \textcolor{red}{$\beta$}, \textcolor{red}{$\gamma$}), the authors discuss how LLMs, despite their numerous advantageous applications, can be employed by cybercriminals for creating destructive payloads and tools. Further, the research methodically generated implementable code for the top-10 MITRE Techniques prevalent in 2022 using ChatGPT and conducted a comparative analysis of its performance with Gemini (formerly known as Bard). 

In \cite{mozes2023use} (\textcolor{red}{*}), the authors discuss the general misuse of LLMs for various applications, not specifically for malware. However, it identifies malware generation as a potential ruinous use case. This paper serves as a primer for the identification of the vulnerabilities of LLMs, as well as mitigating factors for the prevention of misuse. 

In \cite{koide2023detecting} (\textcolor{red}{$\epsilon$}, \textcolor{red}{$\zeta$}), the authors focus on using ChatGPT for detecting phishing sites on the internet. In particular, they propose a new system that uses a web crawler to gather information from websites, generating prompts for LLMs based on the crawled data and retrieving detection results from the responses generated by the LLMs.  

In \cite{kshetri2023cybercrime} (\textcolor{red}{$\alpha$}), the author pens a short paper in which the privacy and security implications of LLMs are explored. In particular, the paper investigates the potential risks posed by LLMs being used by malicious actors and examines the behaviors and practices of developers, operators, and users of these models from privacy and security perspectives.

In \cite{ali2023huntgpt} (\textcolor{red}{$\epsilon$}, \textcolor{red}{$\zeta$}),  the researchers strive to provide cybersecurity researchers and teams with easy-to-use and congenital model interaction to assess and respond to threats effectively. Specifically, they introduce a new framework, built on ChatGTP, that combines the power of ML and LLMs with XAI frameworks to enhance interpretability. The model is benchmarked and tested for accuracy using the KDD99 dataset \cite{olusola2010analysis,ozgur2016review} and the Certified Information Security Manager (CISM) examinations. 

In \cite{bethany2024large} and \cite{heiding2024devising} (\textcolor{red}{$\alpha$}, \textcolor{red}{$\beta$}, \textcolor{red}{$\gamma$}), the authors explore the usage of LLMs in the context of phishing attacks. Overall, these research pieces contribute to the understanding of large-scale attacks and how ML can be used for use cases involving LLM-generated content like phishing emails.  

In \cite{botacin2023gpthreats} (\textcolor{red}{$\alpha$}), the author addresses concerns regarding the negative impact of advanced textual models, such as assisting attackers in the creation of malware. The goal of the research is to investigate whether current large textual models, represented by GPT-3, can be used for generating malware and, if so, to understand how attackers could leverage these models for such purposes. This research provides an appraisal of coding strategies for malware generation and the evaluation of malware variants produced by GPT-3.

Fig. \ref{fig3} illustrates a comprehensive Venn diagram classification of the literature on the usage of LLMs for malware detection. Each section of the diagram represents distinct aspects of research in this domain, encompassing various methodologies and applications. Notably, the diagram highlights the interconnected nature of these studies, with many papers contributing to multiple areas of knowledge. For instance, references such as \cite{pa2023attacker}, \cite{fang2024llm}, and \cite{kshetri2023cybercrime} showcase the dual-use risks of LLMs in generating both beneficial and harmful content. The clustering of references like \cite{yao2024survey}, \cite{shayegani2023survey}, and \cite{motlagh2024large} underscores the importance of addressing security vulnerabilities and adversarial attacks on LLMs. Additionally, the overlaps emphasize the multidimensional challenges and innovative solutions proposed by researchers, such as the usage of LLMs for real-time code analysis and trend identification. This visual taxonomy not only organizes existing research but also identifies gaps for future exploration, particularly in the areas of LLM-based malware generation and sophisticated evasion techniques. Overall, the figure serves as a roadmap for security researchers, guiding them through the complexities and interdependencies within this evolving field.

While we have tried to be as exhaustive in our taxonomy and coverage, we do not claim to have a review of every paper in the field, but at least some of the contemporary and influential works. Other resources are contained therein for further investigation.

We infer from the Venn diagram in Fig. \ref{fig3} that material on the pure weaponization of LLMs (\textcolor{red}{$\beta$}) and LLM-based distribution (\textcolor{red}{$\gamma$}) of malware is non-existent. This is expected as having such information in the public domain would be disastrous. Additionally, we observe that this leaves many research gaps in the security domain, such as the design of algorithms and procedures that simultaneously use LLMs to generate and disseminate malware (\textcolor{red}{$\alpha\cap\gamma$}), creating harmful content using LLMs and then weaponizing it (\textcolor{red}{$\alpha\cap\beta$}), and the design of new models to weaponize LLMs (\textcolor{red}{$\beta\cap\zeta$}). Further research is required in these domains, with a specific emphasis on mitigating the risks associated with turning LLMs into weapons of militarization and cyberwarfare -- Weapons of \textit{large} destruction. Additionally, we observe that a gap in the literature exists in the examination of LLM-specific malware generation and distribution; most of the literature is centered around ChatGPT or Gemini -- This presents an opportunity to examine how other LLMs, mentioned in Tab. \ref{tab1}, fair. 

\begin{figure*}[htpb]
\centering
\includegraphics[width=1\linewidth]{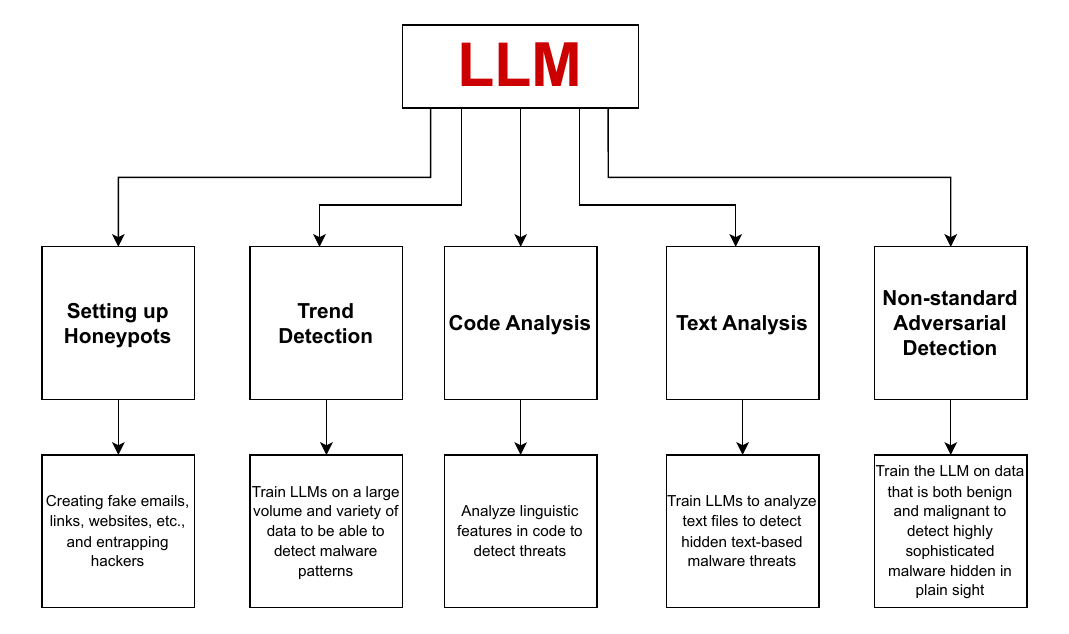}
\caption{Hierarchical classification of various methods through which LLMs can be employed to detect malware. The central node represents the core LLM technology, while the primary branches emanating from it illustrate specific techniques and applications. The secondary branches provide discussion points of their respective parent nodes.}
\label{fig30}
\end{figure*}

\section{The Potential Use of LLMs for Protecting Against Malware -- Problem Formulation and Performance Metrics}\label{problem formulation}

In this section, we describe various malware attack types and discuss how LLMs can be used as a contrivance in their prevention and spread. The caveat here is that while we acknowledge that malware is ``malicious software'' created by the software itself, LLMs can be used to detect patterns in software code, for example, to flag a piece of code as potential malware.

There exists a profusion of methods and schemes to help fight against malware; for example, antivirus software is amongst the most ubiquitous. However, in the age of Generative AI (GenAI), LLMs can be used to gain a competitive advantage and stop a malware attack before or even while it is taking place. Specifically, LLMs can be used in the following manner:
\begin{enumerate}
\item \textbf{Malware Honeypots:} LLMs can generate realistically deceptive content, which includes phishing emails, fake websites, and documents. These can then be used as bait to lure attackers into honeypots. The interaction with these honeypots allows for the collection of data that can be used to enhance LLM training, and thereby improve threat detection capabilities. The engagement of the attackers with the malware honeypots can be calculated by simply counting all the interactions of each attacker with each of the malware honeypots. Mathematically,
\begin{equation}
E=\sum_{i,j}I_{i,j}
\end{equation}
where $I_{i,j}$ is the number of interactions of attacker $i$ with honeypot $j$. 
\item \textbf{Identification of Text-based Threats:} LLMs can be trained to identify malicious code embedded within seemingly benign text files. For example, malware might use obfuscated code within a text-based file to bypass traditional security measures. The effectiveness of the LLM in detecting text-based threats can be evaluated as follows: Firstly, given a text-based file, determine the probability of a threat by building a binary classification model, and thereafter perform a characterization according
\begin{equation}\label{probability eqn}
\mathbb{P}(\text{threat}|\mathbf{x})=\sigma\left(\mathbf{w}^{T}\mathbf{x}+b\right),
\end{equation}
where $\text{threat}=\left\{0,1\right\}$ is a binary variable with $0$ indicating a non-threat, and $1$ indicating a threat, $\mathbf{w},b$ are the model's weights and bias term, $\sigma$ is the sigmoid function, and $\mathbf{x}$ are the model features. Thereafter, all threats can simply be summed up to give a threat score based on the number of predicted matches. Mathematically,
\begin{equation}
S_{\text{threat}}=\sum_{i}p(i),
\end{equation}
where $p$ is a predicted to actual threat match. 
\item \textbf{Analysis of Code to Detect Nefarious Intent:} LLMs can analyze the structure and semantics of code to detect potential malicious intent, even if the code is obfuscated or written in a non-standard way. The LLM can calculate a risk score by analyzing the code in its entirety or a partial snippet. Based on this risk score it can determine whether the code contains harmful objects. Mathematically, this risk score for a code $c$ can be calculated as
\begin{equation*}
R(c)\sum_{n\in T(c)}\Gamma(n)y(n),
\end{equation*}
where $T(c)$ is the abstract syntax tree, $\Gamma: T\to\left[0,1\right]$ such that it is a dictionary mapping from the abstract syntax tree to a numerical value that attempts to assign a weight to common keywords used in malicious code, and $y(n)$ is the output from a classification model that maps the features to a risk score, analogous to Eqn. \eqref{probability eqn}.
\item \textbf{Malware Trend Identification:} LLMs can process large datasets of past malware to identify trends and predict future attacks. This involves clustering similar malware samples and analyzing their evolution over time. One such way of identifying trends on a large corpus of malware data is metric-based clustering. For some latent features $\mathbf{x}_{i}$ and $\mathbf{x}_{j}$ representing the $i^{\text{th}}$ and $j^{\text{th}}$ instances respectively, the metric can be computed using cosine similarity, similar to NLP, 
\begin{equation}
d(\mathbf{x}_{i},\mathbf{x}_{j})=1-\frac{\mathbf{x}_{i}^{T}\mathbf{x}_{j}}{||\mathbf{x}_{i}||_{2}\;||\mathbf{x}_{j}||_{2}}.    
\end{equation}
where $||\cdot||_{2}$ is the Euclidean distance. A distance $d\to1$ indicates a perfect match for malware, and $d\to-1$ indicates that it is not malware.  The time evolution component can be tracked by fitting some Markovian temporal model, say $\mathbf{x}_{i+1}=f(\mathbf{x}_{i},t|\theta)$ at time step $t$ and parameters $\theta$. 
\item \textbf{Non-standard Disguised Malware:} As adversaries become more sophisticated in disguising malware, LLMs can be trained on adversarial examples to enhance detection capabilities. This involves generating adversarial samples and refining the model to improve its robustness. For example, GANs or VAEs can be used to take existing malware data, slightly perturb it, and then train the LLM to minimize the adversarial risk. Mathematically, given some malware sample $x$, perturb it with some $\delta\leq\epsilon$, where $\epsilon\ll 1$ is some small parameter, to create the new malware sample $x'=x+\delta$. Then, train the LLM to solve the optimization problem
\begin{equation} \label{non-standard eqn}
\underset{\theta}{\min}\;\mathbb{E}_{\left(\mathbf{x},y\right)\sim\mathcal{D}}\left\{\underset{\delta\leq\epsilon}{\max}\;J[y,\overbrace{f(x+\delta,\theta)}^{\hat{y}}]\right\},
\end{equation}
where $f$ is the predictor, $J$ is the loss function, $\theta$ are the model parameters, and $\mathcal{D}$ is the training data. The solution to this problem exemplifies how an LLM can use existing malware training data to learn new threats via small perturbations. 
\end{enumerate}

\begin{figure*}[htpb]
\centering
\includegraphics[width=1\linewidth]{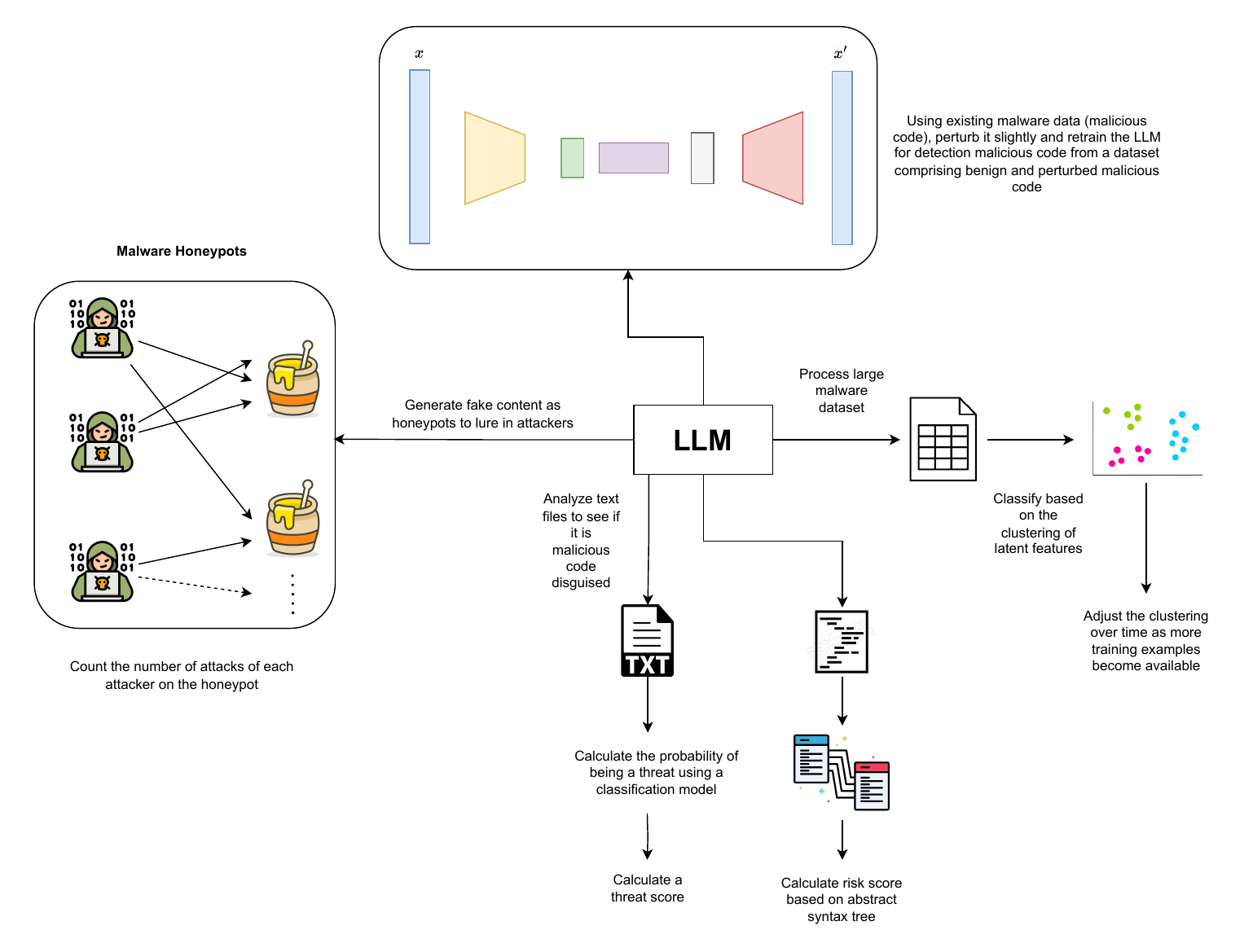}
\caption{Graphical depiction of the mechanics of the proposed metric framework. The diagrams capture the five ways in which LLMs can be used for malware detection, involving simple counting in the case of malware honeypots and sophisticated optimization routines in the case of using existing malware code data to train for the detection of new threats.}
\label{fig20}
\end{figure*}

In Fig. \ref{fig30}, we provide a pictorial representation of the various use cases for which LLMs can be used for protecting against malware, and in Fig. \ref{fig20}, we provide a detailed graphic of the various calculations of the metrics to demonstrate proof of concept.  

\section{Guiding Principles and Risk Mitigation Framework} \label{guidelines}

\subsection{Guiding Principles}
Based upon the literature review above in Sec. \ref{literature review}, in this section, we propose the following guiding principles for using LLMs to detect malware. These are as follows: 
\begin{enumerate}
\item \textbf{Train the Model on a Large Corpora:} The LLM should be trained on a vast amount of data from different sources so as to generalize to out-of-distribution use cases and realistically identify threats. 
\item \textbf{Identify Irregularities and Warning Signs:} Train malware-specific LLMs on data that is composed of both malicious and congenial data that is both code (from different languages having different syntax) and text (documents, emails, etc.). The model should look out for linguistic traits that deviate from normalcy and, therefore, are indicative of maleficent intent.
\item \textbf{Refine Threat Detection:} Continuously improve threat detection capabilities by retraining models on a regular basis. Bearing in mind that LLMs have parameters in the order of $10^{9}-10^{12}$, and are growing, retraining strategies like partial binarized training and compressed-weight training schemes.
\item \textbf{Adopt a Human-centric Approach:} The model should be trained, at least partially, using approaches like Reinforcement Learning from Human Feedback (RLHF) in order to align the intelligence of the LLM with that of human preferences in malware detection. Complete reliance and autonomy cannot be placed on either human beings or computers themselves. Thus, such approaches strike a balance in order to be effective in malware detection. Ultimately, a human expert should have the final say as to whether a threat is legitimate or not; the LLM should be used just as a detection tool.  
\item \textbf{Aggrandize Explainability:} Interpretable AI techniques should be adopted in order to ascertain how LLM models arrived at their conclusion that a particular portion of the code is malware. This promotes trust in the LLM for being able to effectively detect malware threats and establish any biases the model may be inclined towards.
\item \textbf{Performance Monitoring and Model Fine-tuning:} Monitor the performance of LLM-based malware detection systems and adjust the training data and model parameters as needed in order to optimize the precision and accuracy and reduce the false positives. In addition, stay adept with the latest developments in LLM technologies, and incorporate these advancements into the existing models.
\item \textbf{Integrate with Existing Security Infrastructure and Tools:} The LLM malware detection system should be used concurrently with antivirus software, intrusion detection systems, and sandboxing.  
\end{enumerate}
The principles laid out address the concerns and lacuna identified in \cite{motlagh2024large}, and points \textcircled{6} and \textcircled{7} address the need for accommodative cybersecurity strategies, as identified in \cite{vasconcelos2023llama}.  

\begin{figure}[htpb]
\centering
\includegraphics[width=1\linewidth]{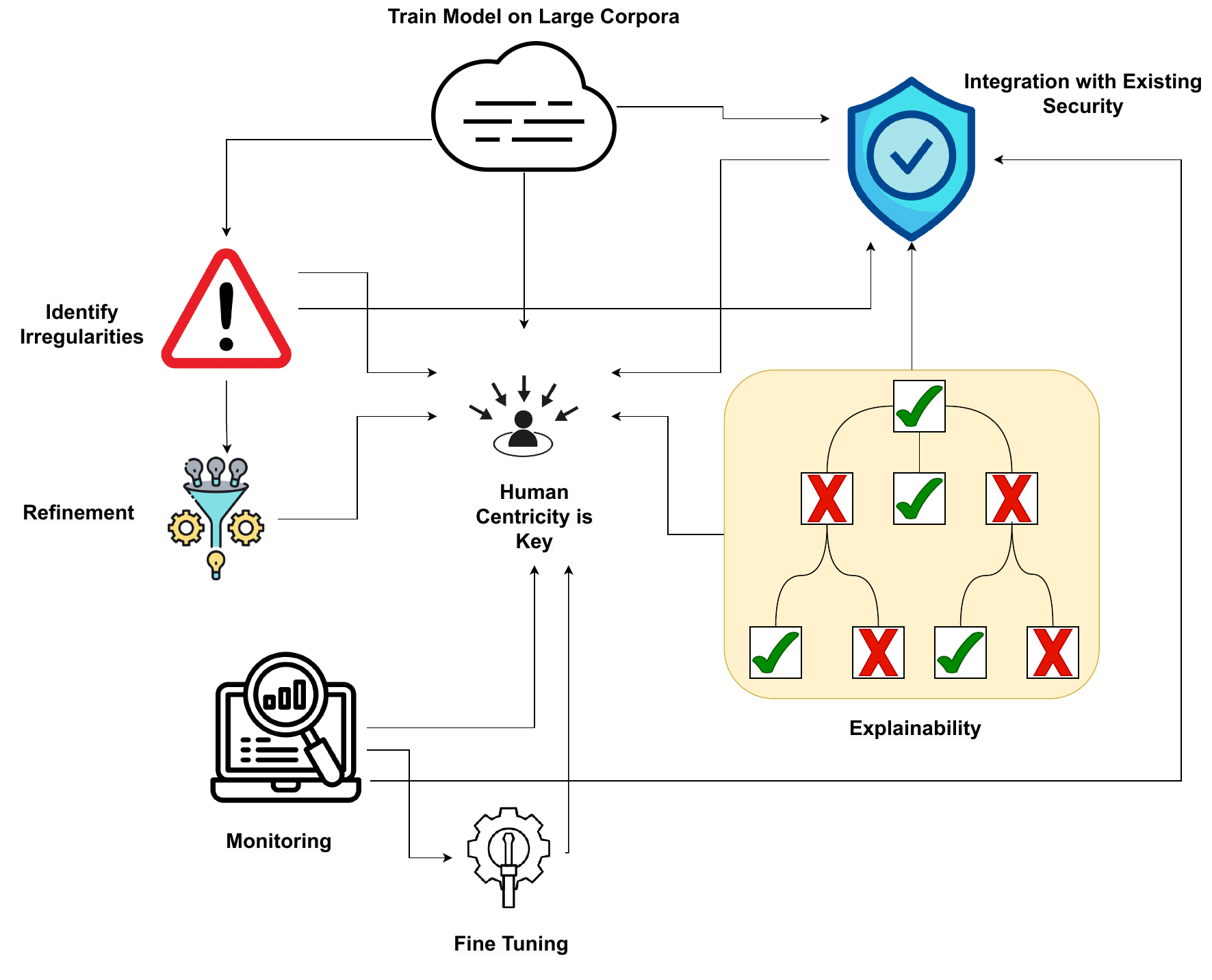}
\caption{A graphical representation of the key guiding principles for effectively utilizing LLMs in malware detection. The diagram outlines a systematic approach, emphasizing the importance of data quality, model training, human oversight, and integration with existing security systems.}
\label{fig4}
\end{figure}

In Fig. \ref{fig4}, we provide a graphical summary of the various guiding principles. It is out belief that these principles cover the various use cases on how to leverage LLMs effectively for preventing risks associated with malware. 

Based upon the guiding principles, we believe that they serve as a beacon for the compliance of organizations' cybersecurity departments to aim for and as a shepherd to guide the conformance of industry and governmental standards. In addition, we believe that these principles are modular, and can be tailored to a particular region, country, industry, or company in conjunction with other standards and regulations in the Age of Big Data such as the General Data Protection Regulation (GDPR) in the EU \cite{voigt2017eu}, the California Consumer Protection Act (CCPA) in the state of California, USA \cite{pardau2018california}, or the Protection of Personal Information Act (POPIA) in South Africa \cite{adams2021popia}. 

\subsection{Risk Mitigation Strategies for Using LLMs to Prevent Malware} \label{risk mitigation}
In this section, we propose a set of risk assuagement plans of action for detecting and preventing malware that is generated using LLMs:
\begin{enumerate}
\item \textbf{Train the LLM Model on Confounded Data Samples:} In order to make the LLM aware of different malware use cases like polymorphisms, encryption, and packing, the LLM should be trained on different samples in order to see through disguises. In addition, LLMs should focus not only on organic data but also on synthetic data from generative models in order to strengthen their ability to generalize.  
\item \textbf{Scan Code in Real-time:} Of course, this is not possible with large LLMs; however, use lightweight LLMs to perform real-time scans on code for early detection of malware. 
\item \textbf{Generate Test Cases from Targeted Sandboxing:} Generate targeted test cases that can be used to trigger the LLM's ability to flag malware in code, thereby allowing for a more comprehensive analysis and creating a dynamic feedback loop.
\item \textbf{Heighten Security and Privacy through Federated Training:} In order to increase the diversity of the training data and thereby its ability to detect differing malware types, consider training the model using federated learning and then generate a global model. This has the additional benefit of not training the model all on one machine, which makes it more agile. 
\item \textbf{Continuous Learning:} Keep the LLM up-to-date with the latest developments in threats and security. This will ensure that it is effective in preventing new types of malware. 
\end{enumerate}

Point \textcircled{5} specifically addresses the need for novel approaches identified in \cite{kang2023exploiting}. 

\begin{figure}[htpb]
\centering
\includegraphics[width=0.9\linewidth]{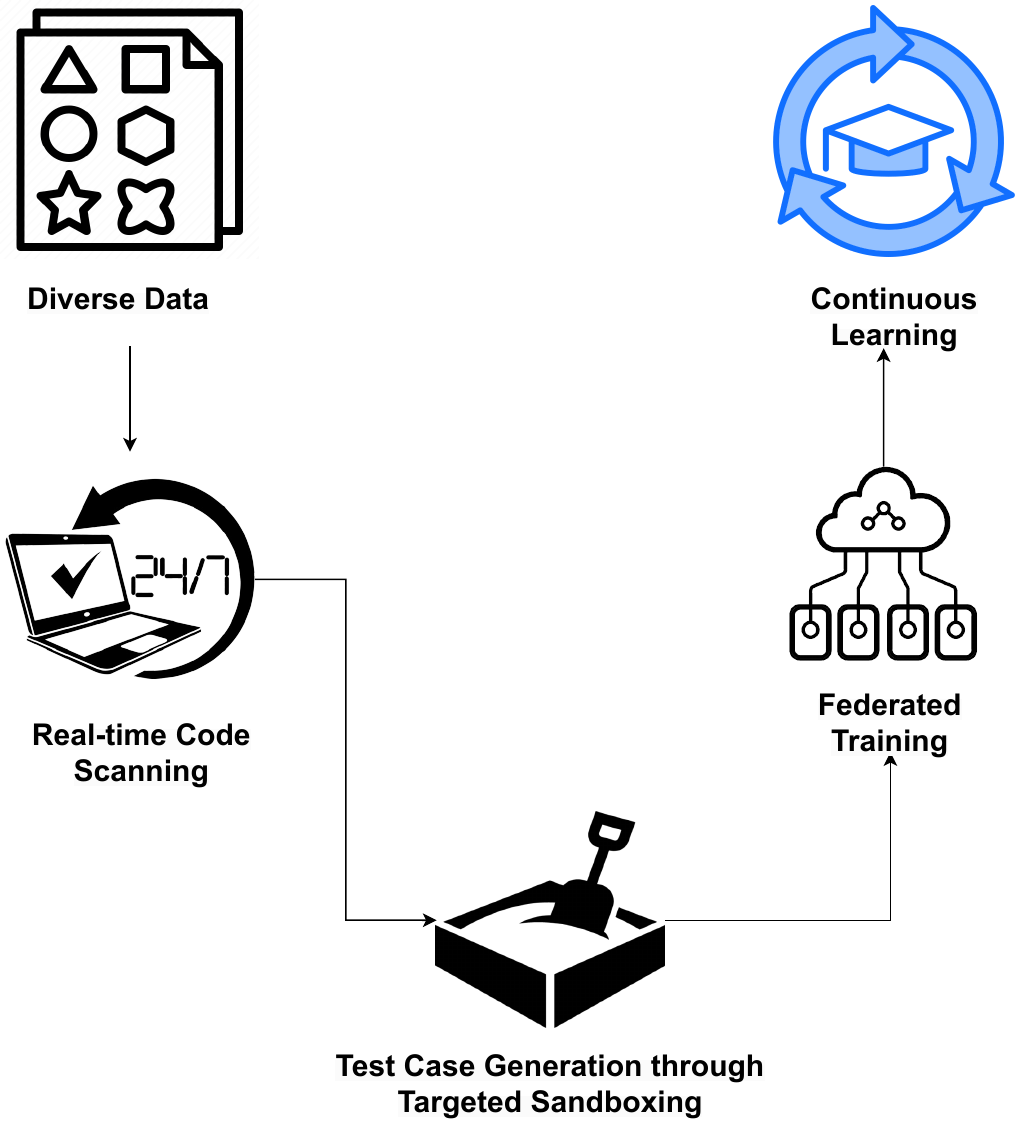}
\caption{A graphical rendition of the proposed risk mitigation strategy for utilizing LLMs in malware detection. The diagram outlines a multi-faceted approach that encompasses data diversity, continuous learning, real-time code scanning, targeted sandboxing, and federated training.}
\label{fig5}
\end{figure}

In Fig. \ref{fig5}, we provide a graphical summary of the proposed risk mitigation strategies.

\section{Performance Evaluation}\label{performance evaluation}
In this section, we create conjectural scenarios that have fictitious data in order to demonstrate how the proposed metrics in Sec. \ref{problem formulation} work. Specifically, we address each metric by providing situation-relevant data that could be collected in a real-world use case. 

\begin{enumerate}
\item \textbf{Malware Honeypots:} Consider the scenario of one attacker that interacts 4 times with the first malware honeypot and 3 times with the second. These interactions could be phishing emails, fake account details, etc., designed to lure the attacker in. Then, the effectiveness is
\begin{equation*}
E=I_{1,1}+I_{1,2}=4+3=7\;\text{units of engagement}. 
\end{equation*}
Metrics can then be tracked over time to form time series plots and be monitored. 
\item \textbf{Identification of Text-based Threats:} Consider the situation where a text-based file had the keywords \texttt{eval} and \texttt{exec}. Given the simplistic model with parameters $\mathbf{w}=\left[0.30,0.72\right]^{T}, 
b=0$, and features for this use case $\mathbf{x}=\left[1.30,2.25\right]^{T}$. Then, the probability it is a text-based threat equals
\begin{equation*}
\mathbb{P}=\sigma\left(\begin{bmatrix}
0.30 &0.72    
\end{bmatrix}
\begin{bmatrix}
1.30 \\
2.25
\end{bmatrix}
+0
\right)\approx 0.8818.
\end{equation*}
So, it is reasonable to assume that there is a high chance that this is a text-based threat. Thereafter suppose that 100 such predictions resulted in these text files being flagged as threats, then the threat score is $S=100$. Internal business rules can be decided upon as to what action needs to be taken. 
\item \textbf{Analysis of Code to Detect Nefarious Intent:} Consider the abstract syntax tree comprising three nodes: $T(c)=\left\{n_{1}-\texttt{eval}, n_{2}=\texttt{exec}, n_{3}=\texttt{for}\right\}$, and corresponding model outputs $y(n)=\left\{0.82,0.94,0.12\right\}$. Using the mapping
\begin{equation*}
\Gamma(n)=
\begin{cases}
0.91,\quad\text{if}\;n=\texttt{eval}, \\
0.83,\quad\text{if}\;n=\texttt{exec}, \\
0.07,\quad\text{if}\;n=\texttt{for}, \\
0.89,\quad\text{if}\;n=\texttt{delete}, \\
\quad\vdots
\end{cases}
\end{equation*}
the risk score is
\begin{align*}
R(c)=&\;\left(0.91\times 0.82\right)+\left(0.83\times 0.94\right)+\left(0.07\times 0.12\right) \\
\approx&\;1.54. 
\end{align*}
It is possible to conclude that values within this range result in code that has the potential to have malicious intent. 
\item \textbf{Malware Trend Identification:} Consider the samples with latent features $\mathbf{x}_{1}=\left[0.23,0.42\right]^{T}$ and $\mathbf{x}_{2}=\left[0.11,0.67\right]^{T}$, where $\mathbf{x}_{1}$ are the features of a cluster in the training data, and $\mathbf{x}_{2}$ are the latent features of potential malware whose trend we are trying to identify. Then, the metric is given by
\begin{align*}
d(\mathbf{x}_{1},\mathbf{x}_{2})=&\;1-\frac{\left(0.23\times0.11\right)+\left(0.42\times0.67\right)}{\sqrt{0.23^{2}+0.42^{2}}\times\sqrt{0.11^{2}+0.67^{2}}} \\
\approx&\;-0.5213.
\end{align*}
This value indicates a low potential for malware based on clustering.

\item \textbf{Non-standard Disguised Malware:} Consider the LLM performing the binary classification problem on a perturbed sample to classify it as malware or not malware. For this task, we use the binary cross-entropy loss function, the model $f(x;\theta)=\theta_{1}x+\theta_{2}$, and hyperparameters $\epsilon=0.1, \delta=0.02, \theta_{1}=0.5, \theta_{2}=-1$, and data in Tab. \ref{tab100}.
\begin{table}[!h]
\centering
\begin{tabular}{ccccc}
\hline 
$x$ &$y$ &$x'=x+\delta$ &$f=\hat{y}=\theta_{1}x'+\theta_{2}$ &Adjusted $\hat{y}=\text{clip}_{\hat{y}}\left(\varepsilon,1-\varepsilon\right)$ \\
\hline 
1 &0 &1.02 &$-0.49$ &$\varepsilon$ \\
2 &1 &2.02 &$0.01$ &$0.01$ \\
3 &0 &3.02 &$0.51$ &$0.51$ \\
4 &1 &4.02 &$1.01$ &$1-\varepsilon$ \\
5 &0 &5.02 &$1.51$ &$1-\varepsilon$ \\
\hline
\end{tabular}
\caption{Sample data for demonstrating Eqn. \ref{non-standard eqn}.}
\label{tab100}
\end{table}
In Tab. \ref{tab100}, we adjust the $\hat{y}$ values between the range $\left[0,1\right]$ by clipping them to the range $\left[\varepsilon,1-\varepsilon\right]$, with $\varepsilon$ chosen to be $10^{-15}$.  

The loss function is then
\begin{align*}
J=&\;-\frac{1}{n}\sum_{i=1}^{n}\left[y_{i}\log\left(\hat{y}_{i}\right)+\left(1-y_{i}\right)\log\left(1-\hat{y}_{i}\right)\right] \\
=&\;-\frac{1}{5}\left[2\log\left(1-\varepsilon\right)+\log\varepsilon+\log\left(0.01\times0.49\right)\right] \\
\approx&\;7.97146.
\end{align*}
\end{enumerate}
As the model training reduces this loss function, the LLM becomes better at identifying such disguised malware.


\section{Conclusion} \label{conclusion}
In this paper, a comprehensive literature review was presented, categorized, and discussed, in sufficient detail, on the use of LLMs to both generate malware and also to detect malware from code. The paper then provides comprehensive risk mitigation strategies and defines a set of performance factors that were used to evaluate the efficacy of the proposed strategies. In particular, the formulated metrics tackled five scenarios: Honeypots, text-based threats, code-based threats, identifying trends in existing malicious software, and using existing malicious software to create new code to train the LLM on. By creating conjectural situations with data, we were able to demonstrate the utility of these metrics. In addition, we suggested a set of general risk mitigation principles on the usage of LLMs for preventing malware and guiding principles on this matter. 

\begin{figure*}[htpb]
\centering
\includegraphics[width=1.0\linewidth]{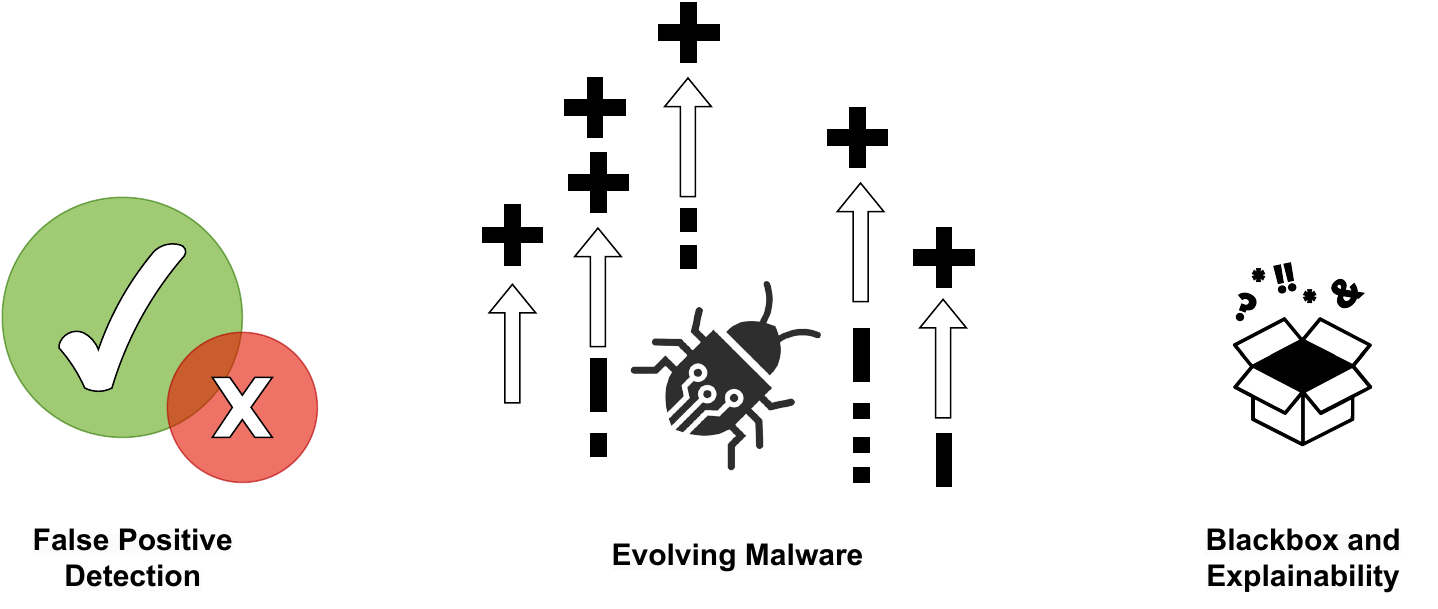}
\caption{The key factors that can influence the effectiveness of LLMs in detecting and preventing various types of malware attacks. The checkmark (\checkmark) in the green category and cross ($\times$) in the red category within a Venn diagram symbolize the challenge of balancing detection accuracy with minimizing false positives, the upward-pointing arrows and a stylized bug icon represents the continuous evolution of malware threats, and the punctuation marks and a partially open box symbolize the inherent black-box nature of LLMs and the need for greater explainability.}
\label{fig6}
\end{figure*}

However, while it holds immense potential, it should be noted that LLMs are not a silver bullet that can prevent or stop every genus of malware attack that exists. This is owing to the following three factors, which are graphically depicted in Fig. \ref{fig6}. First, the detection of false positives. As with any ML algorithm, the probability of detecting type 1 errors, where an affirmative for a malware attack when there is none exists. This is owing to the complexities of how hackers -- who are human beings -- write out and structure their code. Human beings have a tendency to be sloppy and use irrational variable assignments and sloppy code to bypass intrusion detection. Second, the ever-changing nature of malware. Malware does not remain static! Hackers are constantly changing around the code in order to make it detection-resistant. Thus, LLMs need to constantly be trained on new data to stay ahead of the curve and be resilient against such attacks. Finally, the explainability factor where LLMs have complex architectures, and therefore, it is very difficult to nearly impossible to explain the reasoning behind why a certain decision is made, for example, the flagging of a program as being malware. This is not only a problem with LLMs but with AI models in general; thus, this presents an argument for why we cannot be solely reliant on LLMs to definitively be our de facto malware detection systems. 

Auxiliary future areas of exploration would be on leveraging LLMs for detecting malware on 4IR and 5IR devices, say on quantum computers -- \textit{quantum malware} -- whereby there exists potential exploitation of quantum mechanical properties such as superposition and entanglement, or even the manipulation of the quantum data which exist as superimposed states, addressing the need for next-generation malware detection as highlighted in \cite{madani2023metamorphic}. 


\section*{Declarations}

\begin{itemize}
\item \textbf{Funding:} J.A.K. and M.A.Z. acknowledge that this research is supported by grant number 23070, provided by Zayed University and the government of the UAE.
\item \textbf{Conflict of interest/Competing interests:} The authors declare that there are no conflicts of interest. 
\item \textbf{Ethics approval and consent to participate:} None required.
\item \textbf{Consent for publication:} The authors grant full consent to the journal to publish this article.
\item \textbf{Data availability:} N/A
\item \textbf{Materials availability:} N/A
\item \textbf{Code availability:} N/A  
\item \textbf{Author contribution:} All authors have contributed equally to this research.
\end{itemize}


\end{document}